\documentclass{n}
\usepackage{amsmath,amssymb,graphicx,hyperref}   
\usepackage[pdftex]{color}

\usepackage{lineno}
 
\newcommand{\apj}{Astrophys. J.}
\newcommand{\aap}{Astron. Astrophys.}
\newcommand{\apjl}{Astrophys. J. Lett.}
\newcommand{\apjs}{Astrophys. J. Suppl. Ser.}
\newcommand{\mnras}{Mon. Not. R. Astron. Soc.}

\newcommand{\nat}{Nature}

\newcommand{\swd}{Schwarzschild }

\newcommand{\kms}{km\,s$^{-1}$}

\title{\large The stellar orbit distribution in present-day galaxies inferred from the CALIFA survey}

\author{Ling Zhu$^1$\thanks{E-mail: lzhu@mpia.de}, Glenn van de Ven$^{1,2}$, Remco van den Bosch$^1$, Hans-Walter Rix$^1$, Mariya Lyubenova$^{3,2}$, Jes\'us Falc\'on-Barroso$^{4,5}$, Marie Martig$^{1,6}$, Shude Mao$^{7,8,9}$, Dandan Xu$^{10}$, Yunpeng Jin$^8$, Aura Obreja$^{11,12}$, Robert J. J. Grand$^{10,13}$, Aaron A. Dutton$^{12}$, Andrea V. Macci$\mathrm{\grave{o}}^{12}$, Facundo A. G\'omez$^{14,15}$, Jakob C. Walcher$^{16}$, Rub\'en Garc\'ia-Benito$^{17}$, Stefano Zibetti$^{18}$, Sebastian F. S\'anchez$^{19}$
}

\begin{document}

\maketitle
\begin{affiliations}
\footnotesize{
\item Max Planck Institute for Astronomy, K\"onigstuhl 17, 69117 Heidelberg, Germany
\item European Southern Observatory, Karl-Schwarzschild-Str. 2, 85748 Garching b. M\"unchen, Germany
\item Kapteyn Astronomical Institute, University of Groningen, Postbus 800, NL-9700 AV Groningen, the Netherlands
\item Instituto de Astrof\'isica de Canarias (IAC), E-38205 La Laguna, Tenerife, Spain
\item Universidad de La Laguna, Dpto. Astrof\'isica, E-38206 La Laguna, Tenerife, Spain
\item Astrophysics Research Institute, Liverpool John Moores University, 146 Brownlow Hill, Liverpool L3 5RF, UK
\item Physics Department and Tsinghua Centre for Astrophysics, Tsinghua University, Beijing 100084, China
\item National Astronomical Observatories, Chinese Academy of Sciences, 20A Datun Road, Chaoyang District, Beijing 100012, China
\item Jodrell Bank Centre for Astrophysics, School of Physics and Astronomy, The University of Manchester, Oxford Road, Manchester M13 9PL, UK
\item Heidelberg Institute for Theoretical Studies, Schloss-Wolfsbrunnenweg 35, D-69118 Heidelberg, Germany
\item Universit\"ats-Sternwarte, Ludwig-Maximilians-Universit\"at M\"unchen, Scheinerstr. 1, D-81679 M\"unchen, Germany
\item New York University Abu Dhabi, PO Box 129188, Saadiyat Island, Abu Dhabi, UAE
\item Zentrum f\"{u}r Astronomie der Universit\"{a}t Heidelberg, Astronomisches Recheninstitut, M\"{o}nchhofstr. 12-14, 69120 Heidelberg, Germany
\item Instituto de Investigaci\'on Multidisciplinar en Ciencia y Tecnolog\'ia, Universidad de La Serena, Ra\'ul Bitr\'an 1305, La Serena, Chile
\item Departamento de F\'isica y Astronom\'ia, Universidad de La Serena, Av. Juan Cisternas 1200 N, La Serena, Chile
\item Leibniz-Institut f\"ur Astrophysik Potsdam (AIP), Germany
\item Instituto de Astrof\'isica de Andaluc\'ia (CSIC), P.O. Box 3004, 18080 Granada, Spain
\item Osservatorio Astrofisico di Arcetri Largo Enrico Fermi 5 I-50125 Firenze, Italy
\item Instituto de Astronom\'ia, Universidad Nacional Auton\'oma de M\'exico, A.P. 70-264, 04510, M\'exico, D.F.
}
\end{affiliations}

\begin{abstract}
Galaxy formation entails the hierarchical assembly of mass, along with the condensation of baryons and the ensuing, self-regulating star formation \cite{MBW2010, White1978}. 
The stars form a collisionless system whose orbit distribution retains dynamical memory that can constrain a galaxy's formation history \cite{Lagos2017b}.
The ordered-rotation dominated orbits with near maximum circularity $\lambda_z \simeq1$ and the random-motion dominated orbits with low circularity $\lambda_z \simeq0$ are called kinematically cold and kinematically hot, respectively.
The fraction of stars on `cold' orbits, compared to the fraction of stars on `hot' orbits, speaks directly to the quiescence or violence of the galaxies' formation histories \cite{Bird2013, Rottgers2014}. 
Here we present such orbit distributions, derived from stellar kinematic maps via orbit-based modelling for a well defined, large sample of 300 nearby galaxies. The sample, drawn from the \emph{CALIFA} survey \cite{CALIFA12}, includes the main morphological galaxy types and spans the total stellar mass range from $10^{8.7}$ to $10^{11.9}$ solar masses. 
Our analysis derives the orbit-circularity distribution as a function of galaxy mass, $p(\lambda_z~|~M_\star)$, and its volume-averaged total distribution, $p(\lambda_z)$. We find that across most of the considered mass range and across morphological types, there are more stars on `warm' orbits defined as $0.25\le \lambda_z \le 0.8$ than on either `cold' or `hot' orbits.
This orbit-based ``Hubble diagram'' provides a benchmark for galaxy formation simulations in a cosmological context.
\end{abstract}

The \emph{CALIFA} survey \cite{CALIFA12} has delivered high-quality stellar kinematic maps for an ensemble of 300 galaxies, complemented by homogeneous $r$-band imaging for all the galaxies from the Sloan Digital Sky Survey (SDSS) DR8 \cite{SDSSdr8}.
This sample of 300 galaxies encompasses the main morphological galaxy types, with a total stellar masses, $M_*$, ranging between $10^{8.7}$ to $10^{11.9}$ solar masses $M_{\odot}$.
\emph{CALIFA}'s selection function is well-defined between $10^{9.7}$ to $10^{11.4}$  $M_{\odot}$ \cite{Walcher2014}, so that sample results within this mass range can be linked to volume-corrected density functions and global averages for galaxies in the present-day universe. 

We construct orbit-superposition Schwarzschild \cite{Schwarzschild1993} models for each galaxy that simultaneously fit the observed surface brightness and stellar kinematics (see Methods). 
In this manner, we find the weights of different orbits that contribute to the best-fitting model.   
We characterize each orbit with two main properties: the time-averaged radius $r$ which represents the size of the orbit, and circularity $\lambda_z\equiv J_z/J_{\mathrm{max}}(E)$, which represents the angular momentum of the orbit around the short $z$-axis normalized by the maximum of a circular orbit with the same binding energy $E$. Circular orbits have $\lambda_z=1$, radial orbits, and more importantly box orbits, have $\lambda_z=0$; counter-rotating orbits have negative $\lambda_z$. 
The resulting probability density of orbit weights, $p(\lambda_z, r)$, is basically a physical description of the 6D phase-space distribution in a galaxy.

\begin{figure}
\centering\includegraphics[width=7cm]{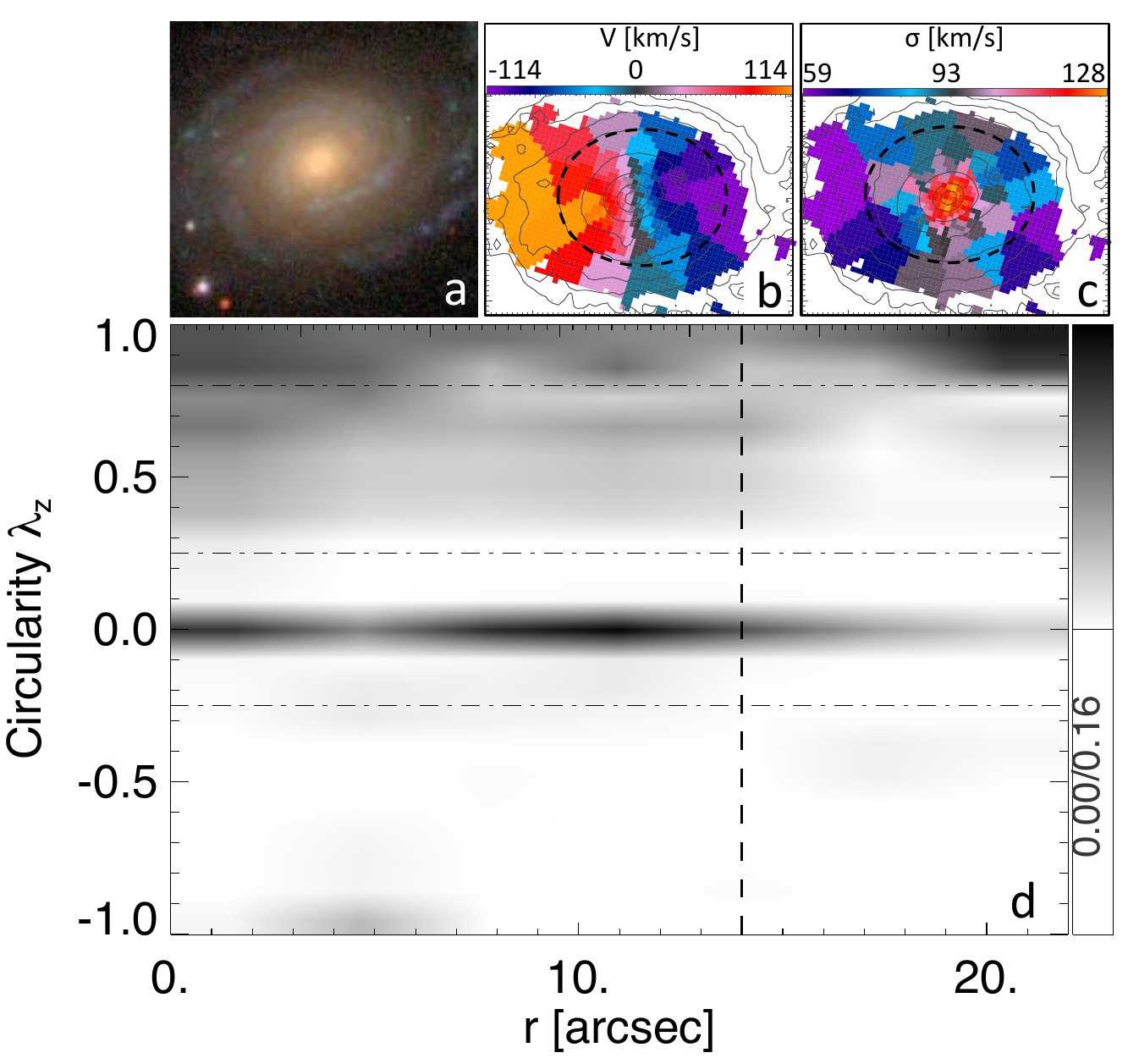}
\caption{ \textbf{Our orbit-based modelling illustrated for galaxy NGC0001.} Panel ({\bf a}) is a SDSS multi-colour image of NGC0001, while panels ({\bf b}) and ({\bf c}) show the CALIFA stellar mean velocity $V$ map and velocity dispersion $\sigma$ map, respectively. Each of the top panels measures 1~arcmin on the side and the dashed ellipse with semi-major axis radius $R_e$ encloses half of the galaxy’s light in the SDSS $r$-band. The orbit-based model that best-fits the SDSS $r$-band image and CALIFA stellar kinematic maps of NGC0001, yields the orbital distribution $p(\lambda_z, r)$ shown in panel ({\bf d}) as function circularity $\lambda_z$ and intrinsic radius $r$. The vertical dashed line denotes the SDSS $r$-band half-light or effective radius $R_e$. Darker color indicates higher probability density as indicated by the grey-scale bar with maximum value chosen such that the sum of all orbital weights is normalised to unity.  }
\label{fig:NGC0001}
\end{figure}

Our orbit-based modeling approach is illustrated in Figure~\ref{fig:NGC0001} for the galaxy NGC0001; it shows the galaxy image and stellar kinematic maps (top) plus the orbit distribution $p(\lambda_z, r)$
of the best-fitting model.  For each of the 300 \emph{CALIFA} galaxies we obtained in this way such an orbit distribution $p(\lambda_z, r)$. Next, integrating $p(\lambda_z, r)$ over all radii $r < R_{\rm e}$ yields the overall orbit circularity distribution, $p(\lambda_z)$, normalized to unity within the effective radius 
$R_{\rm e}$ (defined as the radius which encloses half of the galaxy's light). 

\begin{figure*}
\centering\includegraphics[width=14cm]{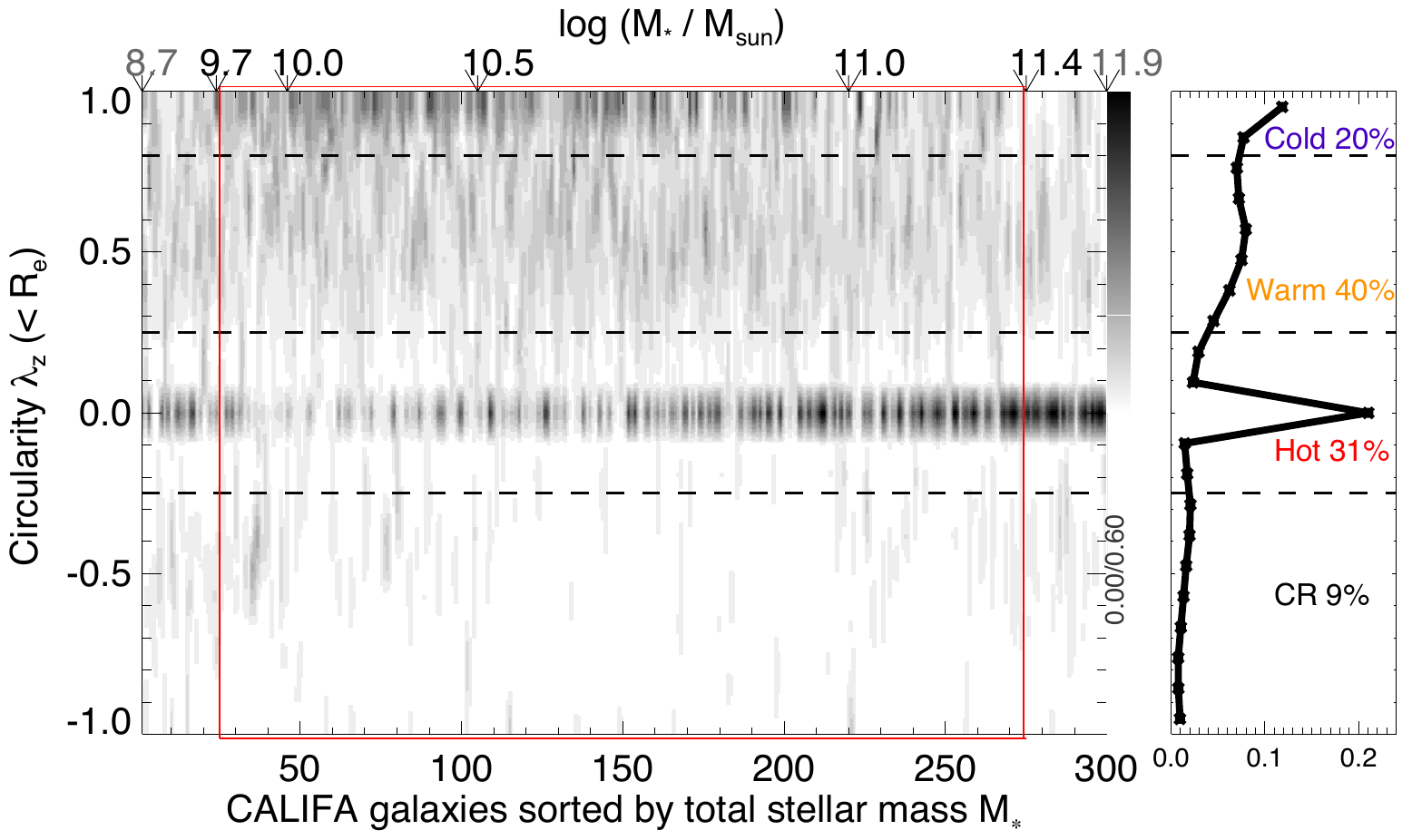}
\caption{ \textbf{The orbit-circularity $\lambda_z$ distribution for each of 300 CALIFA galaxies.} Each thin slice vertically represents the $\lambda_z$ distribution of one galaxy normalized to unity within the half-light- radius Re. Darker color indicates higher probability as illustrated by the grey-scale bar with maximum value chosen such that per galaxy the orbital weights add up to unity over the 21 bins across the range in $\lambda_z$. From left to right, the galaxies are sorted with increasing total stellar mass $M_*$ as indicated at the top. The red box delineates the range of $9.7<\log (M_* /M_{\odot})<11.4$ where the \emph{CALIFA} sample is statistically representative. The right panel shows the volume-corrected average orbit-circularity distribution within this mass range. The right panel shares the $y$ axis of the left panel, while its $x$ axis is probability per bin (with total probability normalized to unity). The cold, warm, hot and counter-rotating (CR) components are divided in $\lambda_z$ indicated by the three horizontal dashed lines.}
\label{fig:orbit300_1Re}
\end{figure*}

Figure~\ref{fig:orbit300_1Re} shows these distributions $p(\lambda_z)$ for the 300 \emph{CALIFA} galaxies, sorted by increasing total stellar mass $M_*$. There are clear overall patterns, but also a great deal of galaxy-by-galaxy variation. We divide the orbits into four broad regimes: (i) cold orbits with $\lambda_z \geq 0.80$ which are close to circular orbits, (ii) warm orbits with $0.25<\lambda_z <0.80$ which are so-called short-axis tube with still a distinct sense of rotation but already considerable random motions, (iii) hot orbits with $|\lambda_z|\leq 0.25$ which are mostly so-called box orbits and a small fraction long-axis tube orbits. 

Of the 300 galaxies, 279 are identified as non-interacting, and of these 269 have kinematic data coverage $R_{\mathrm{max}} > R_{\rm e}$. We further exclude 19 objects that could be biased by dust lanes (see Methods).
Of the remaining 250 galaxies, about half have a bar which is not explicitly modelled. However, our tests with simulated galaxies (see Methods) indicate no significant effect in the recovered orbit distribution. We thus keep 250 galaxies, of which 221 fall within the total stellar mass range $9.7<\log (M_* /M_{\odot})<11.4$ (indicated by the red box in Figure~\ref{fig:orbit300_1Re}), where the \emph{CALIFA} sample is statistically representative. 
We use these 221 galaxies to calculate volume-averaged properties, $f(x)$, for those properties represented by functions, $f_i(x)$, which can be ensemble-averaged:
\begin{equation}
f(x) = \frac{\sum_{i} f_i(x) \times M_{*,i} \times 1/V_i } { \sum_{i} M_{*,i} \times 1/V_i},
\label{eq:ff}
\end{equation}
where $M_{*,i}$ is the total stellar mass and $V_i$ is the volume for each galaxy $i$ \cite{Walcher2014}. If $f_i(x) = p_i(\lambda_z)$, then the resulting $p(\lambda_z)$ is the average orbital distribution of present-day galaxies as shown in the subpanel on the right side of Figure~\ref{fig:orbit300_1Re}. 

We can now characterize the importance (within $R_{\rm e}$) of cold, warm, hot, and counter-rotating orbits in galaxies of different $M_*$. To this end we divide the 250 isolated galaxies with enough data coverage into eight comparably populated mass bins, and then calculate the average orbital distributions in each bin based on Eq.~\ref{eq:ff}. 
Note that the bins with the lowest and highest stellar mass are outside the completeness range of the CALIFA sample even after volume-correction.
The resulting averaged luminosity fraction (SDSS r-band) of the cold, warm, hot and CR components as a function of $M_*$ are shown in Figure~\ref{fig:lzd}.
The values are listed in the Supplementary Table 1 and include assymetric errors bars which, in addition to statistical uncertainties, include systematic biases and systematic uncertainties based on careful calibration against galaxy simulations (see Methods).

We find that the cold component (blue) rarely dominates within $R_{\rm e}$, but is most prevalent among galaxies with total stellar mass $M_* \eqsim 1-2\times 10^{10}\,M_{\odot}$ and decreases for more massive galaxies.
In most galaxies substantially more stars within $R_{\rm e}$ are on warm orbits (orange) that still have distinct angular momentum. 
The hot component (red) rises rapidly with increasing stellar mass and dominates in the most massive galaxies with $M_* > 10^{11}\,M_{\odot}$. 
The counter-rotating (CR) component (black) is roughly constant at $\sim10$\% in all galaxies, except in low-mass $M_* < 10^{10} \, M_{\odot}$ galaxies where the CR component increases seemingly in favor of a decreasing cold component fraction. 

Such stellar orbit distributions have not been derived explicitly for a large sample of galaxies before.  Instead, the ratio of ordered-to-random motion and the flattening of galaxies are two commonly used proxies for the angular momentum of galaxies. To have a more direct comparison with current results, we determine from our models two similar proxies and compare them to the orbital circularity. 

First, we compute the ordered-to-random motion $\langle V_{\phi}/\sigma \rangle_i$ per bin in $\lambda_z$ for each galaxy $i$. Here, $V_{\phi}$ is the intrinsic rotation velocity around the minor axis and $\sigma = \sqrt{\sigma_x^2 + \sigma_y^2 + \sigma_z^2}$ is the intrinsic velocity dispersion.
Taking $f_i(x) = \langle V_{\phi}/\sigma \rangle_i (\lambda_z)$ in equation~(\ref{eq:ff}) yields $\langle V_{\phi}/\sigma \rangle (\lambda_z)$ shown in the top panel of Figure~\ref{fig:zR300}.
As expected, $\langle V_{\phi}/\sigma \rangle$ strongly correlates with $\lambda_z$, and is thus a good indicator of the underlying orbit distribution.
However, the \emph{observed} ordered-to-random motion ratio $\langle V_{\mathrm{los}}/\sigma_{\mathrm{los}} \rangle$ is a projected quantity along an often-unknown line-of-sight viewing angle.
Therefore, the combination with the observed galaxy ellipticity $\epsilon$ is used to constrain the internal dynamics of galaxies \cite{Binney2005}. 
The resulting $(V_{\mathrm{los}}/\sigma_{\mathrm{los}}, \epsilon)$ diagram allows 
slow rotator early-type galaxies to be distinguished from fast rotator galaxies \cite{Cappellari2007}.
Slow rotators are found to be more triaxial and more massive, dominating above $2 \times 10^{11} M_{\odot}$ \cite{Emsellem2011}\cite{Brough2017}. 
This is in agreement with our more direct orbit-based finding of a gradual increase of the hot component with galaxy mass and dominance above $2\times 10^{11} M_{\odot}$.

Second, when considering the flattening of galaxies, it is plausible that stars on cold orbits form thin disks as a consequence of galaxy formation. However, from galaxy dynamics {\it per se} a spherical galaxy could consist of randomly oriented high-angular momentum orbits, that is still dynamically cold with high $\lambda^2_{\mathrm{total}} = \lambda_z^2 + \lambda_x^2 + \lambda_y^2$, but random $\lambda_z$. 
To quantify the relation between flattening and $\lambda_{\mathrm{total}}$, we calculate for each galaxy with luminosity density $\rho$ the geometric flattening 
$\langle q \rangle^2 = (\int\rho z^2 ) / ( \int\rho (x^2 + y^2)/2)$. 
At all galaxy masses, there is a correlation between flattening $\langle q \rangle_i$ and $\lambda_{\mathrm{total}}$. 
We take $f_i(x) = \langle q \rangle_i(\lambda_{\mathrm{total}})$ in Eq.~\ref{eq:ff} to derive an 'universal' mean relation $\langle q \rangle (\lambda_{\mathrm{total}})$ shown in the bottom panel of Figure~\ref{fig:zR300}. 
As expected, the previously defined four components in $\lambda_z$ are consistent with the transition from a flat to a spheroidal distribution.
We have a few percent of long-axis tube orbits in triaxial galaxies, with small but non-zero $\lambda_{\mathrm{total}}$, causing the peak $\langle q \rangle > 1$ at $\lambda_{\mathrm{total}} \sim 0.2$.

Our \swd models thus imply that the cold orbits with high $\lambda_{\mathrm{total}}$ in galaxies always form a highly flattened configuration and only the hot orbits with low $\lambda_{\mathrm{total}}$ form a spheroidal configuration. 
Consequently, the bulge-disk morphology of a galaxy is generally indeed an indicator of the underlying orbit types, but still an inaccurate proxy \cite{Obreja2016} due to the significant scatter of $\langle q \rangle$ as a function of $\lambda_{\mathrm{total}}$ or $\lambda_z$.
Even so, photometric bulge-disk decompositions show that the bulge fraction of spiral galaxies increases with mass above $M_* \sim 10^{10} \,M_{\odot}$ \cite{Weinzirl2009}, which is consistent with the relative increase of the hot component we find. 
From photometric decomposition of edge-on disk galaxies, low-mass spiral galaxies are found to have disks that are thicker than the main thin-disk component in high-mass spiral galaxies \cite{Comeron2012}. This is also consistent with our result that for galaxies with $M_* < 10^{10} \, M_{\odot}$, despite being mainly disk-dominated Sc and Sd galaxies, the cold orbit fraction drops, while the warm orbit fraction remains high and the fraction of counter-rotating orbits increases. 
Our finding that warm orbits constitute the majority within $1\,R_{\rm e}$ for all galaxies, except for the very most massive ellipticals, is a new results and also insight. But it may be a natural  consequence of galaxies growing inside-out, with those stars born earlier having lower angular momentum \cite{Lagos2017, Stinson2013, Bird2013}.  

We find that high-$\lambda_z$ orbits generally form flat and rapidly rotating disks, while low-$\lambda_z$ orbits form near-spherical and slow-rotating spheroids. Our distributions $p(\lambda_z)$ are therefore a (conceptually improved) alternative to geometric and photometric bulge-disk decompositions, or to the qualitative measures of 'galaxy morphology'. In this sense, our results present an orbit-based alternative to the 'Hubble-sequence', characterizing for the first time the internal dynamical structure for a large sample of galaxies.

Given that CALIFA's selection function allows the correction of the sample to volume averages, our results represent an observationally-determined orbit distribution of 'galaxies in the present-day universe'. They lend themselves thus to direct comparison with samples of cosmological simulations of galaxies in a cosmological context\cite{Illustris2014,Eagle2015, Wang2015,Hopkins2014}; $p(\lambda_z)$, averaged within $R_{\rm e}$ can be extracted directly from simulations. 
In this sense, our results open a quantitative way -- and at the same time a
qualitatively new window -- for comparing galaxy simulations to the observed galaxy population in the present-day universe.

\paragraph{References}

\begin{addendum}
\item [Correspondence:] Requests for materials should be addressed to Ling Zhu (lzhu@mpia.de) and/or Glenn van de Ven (gvandeve@eso.org).

 \item This study makes use of the data provided by the Calar Alto Legacy Integral Field Area (CALIFA) survey (http://califa.caha.es). Based on observations collected at the Centro Astron\'omico Hispano Alem\'an (CAHA) at Calar Alto, operated jointly by the Max-Planck Institut f\"ur Astronomie and the Instituto de Astrof\'isica de Andaluc\'ia (CSIC). We thank Arjen van der Wel, Knud Jahnke, Victor Debattista and Morgan Fouesneau for useful discussions. 
GvdV and JF-B acknowledge support from the DAGAL network from the People Programme (Marie Curie Actions) of the European Union’s Seventh Framework Programme FP7/2007- 2013/ under REA grant agreement number PITN-GA-2011-289313. GvdV also acknowledges support from the Sonderforschungsbereich SFB 881 "The Milky Way System" (subprojects A7 and A8) funded by the German Research Foundation, and funding from the European Research Council (ERC) under the European Union's Horizon 2020 research and innovation programme under grant agreement No 724857 (Consolidator Grant 'ArcheoDyn').
This work was supported by the National Science Foundation of China (Grant No. 11333003, 11390372 to SM). AO has been funded by the Deutsche Forschungsgemeinschaft (DFG, German Research Foundation) -- MO 2979/1-1.
JF-B acknowledges support from grant AYA2016-77237-C3-1-P from the Spanish Ministry of Economy and Competitiveness (MINECO).

\item[Authors Contributions:]
Text, figures, and interpretation: LZ, GV, HR, MM, SM, Modelling: LZ, RB, GV; Observational data: JFB, ML, GV, JW, RGB, SZ, SS; Methodology: LZ, DX, YJ, AO, RG, AM, AD, FG.

\item [Supplementary pdf file] Including the tables and figures showing reliability of the methods, and tables of data points showing in Figure 3.

\item[Competing financial interests] The authors declare that they have no
competing financial interests.
\end{addendum}

\newpage

\begin{figure}
\centering\includegraphics[width=8cm]{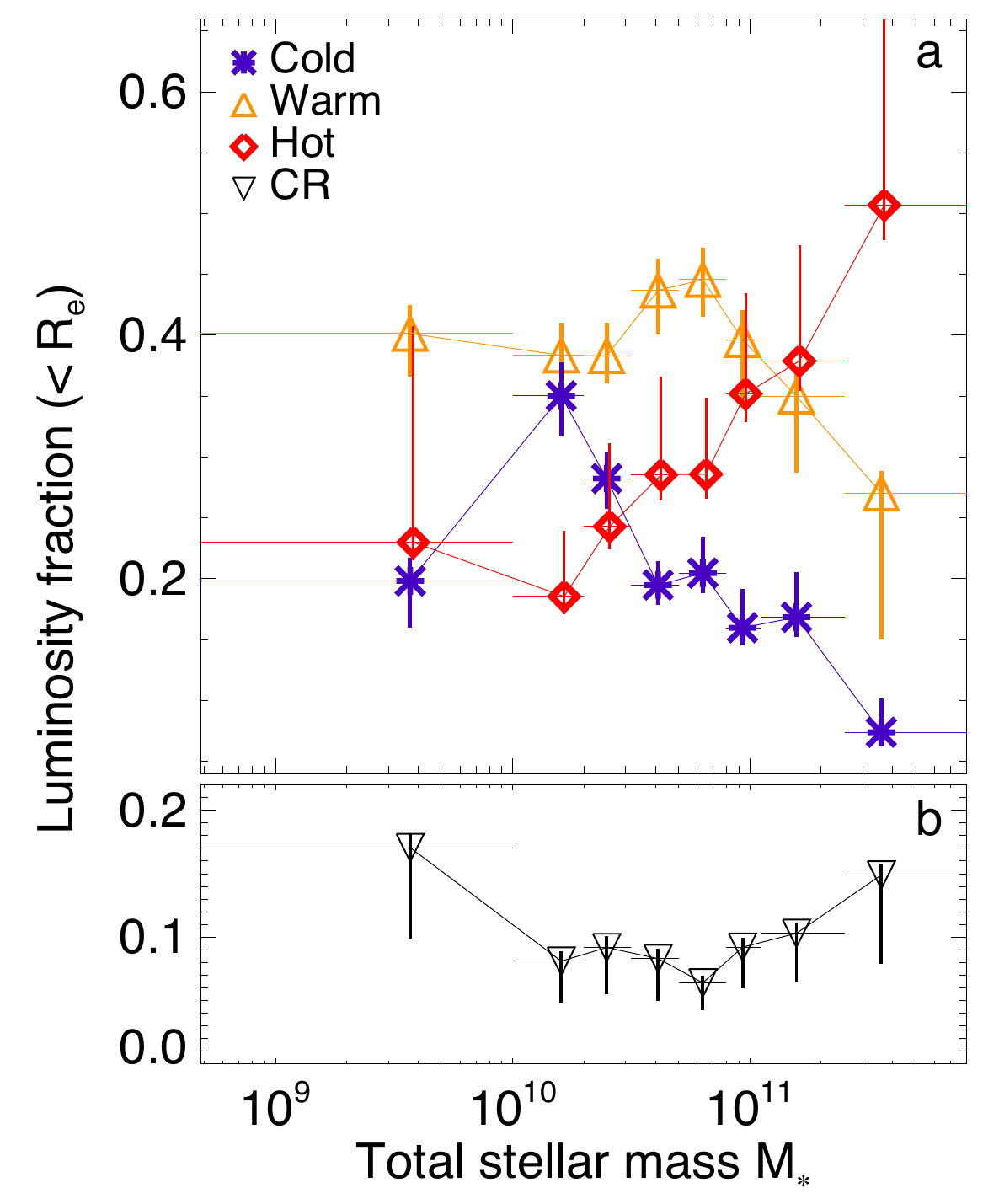}
\caption{ \textbf{The distribution of orbital components as function of galaxy mass.} The CALIFA galaxies are divided into 8 bins in total stellar mass $M_*$ as indicated by the extent of the horizontal error bars. For each bin in $M_*$, the volume-corrected average SDSS $r$-band luminosity fraction within the effective radius $R_e$ is computed for each of four orbital components based on a selection in orbital-circularity $\lambda_z$ as indicated by the horizontal dashed lines in Figure~2. The blue asterisks, orange triangles and red diamonds in panel ({\bf a}) show the resulting fractions for the cold, warm and hot components, respectively, whereas that of the counter-rotating (CR) component is shown with black upside-down triangles separately in panel ({\bf b}) for clarity. The vertical error bars indicate the $1\sigma$ uncertainties, including both statistical uncertainties as well as systematic biases and uncertainties as inferred from tests with simulated galaxies (see Methods).}
\label{fig:lzd}
\end{figure}

\begin{figure}
\centering\includegraphics[width=8cm]{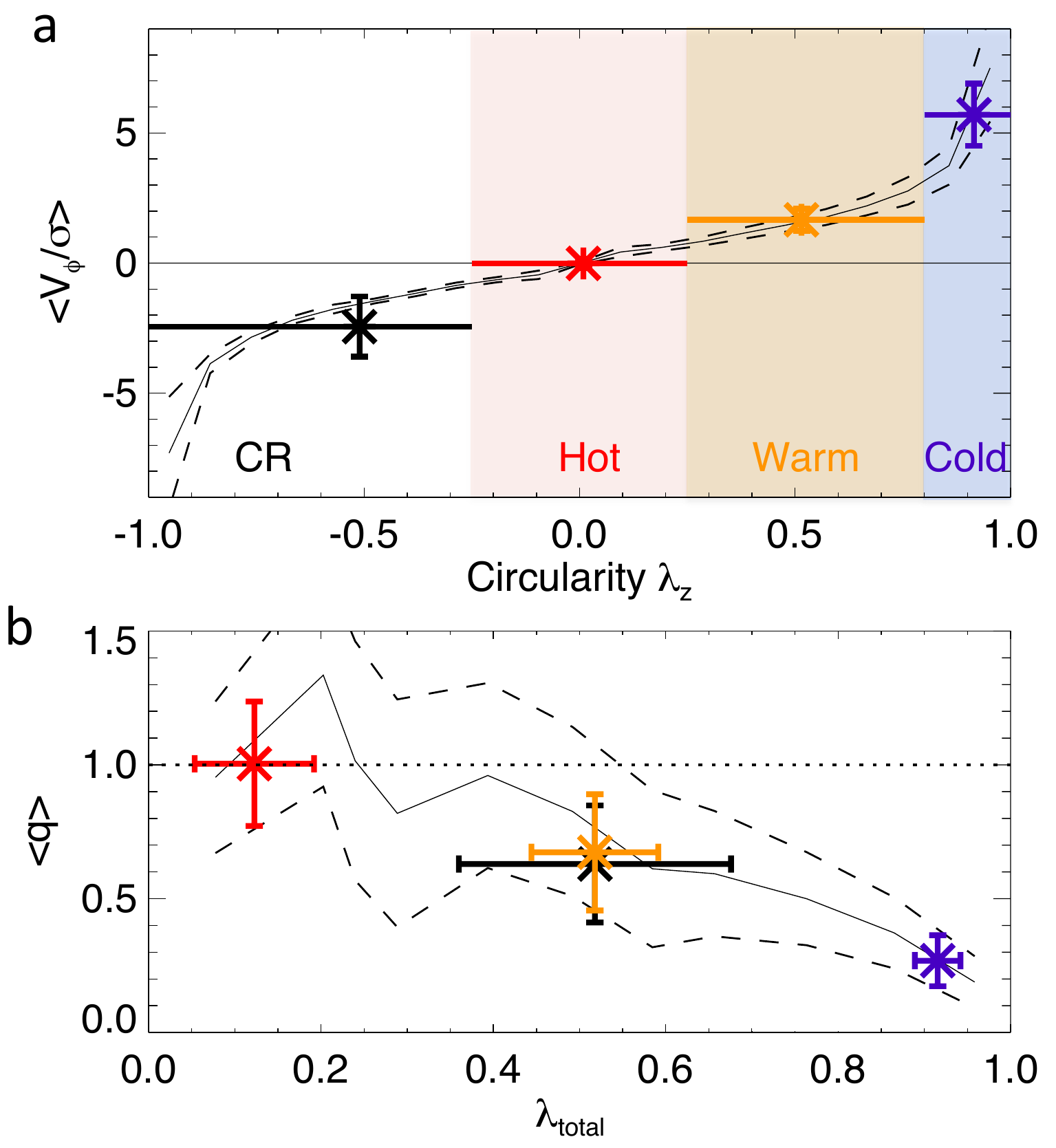}
\caption{\textbf{The relation between intrinsic orbital angular momentum and its two proxies, a galaxy's ordered-to-random motion and geometric flattening.} Panel ({\bf a}) shows the intrinsic ordered-to-random motion $\langle V_{\phi}/\sigma \rangle$ as function of circularity $\lambda_z$. Our division into cold, warm, hot and counter-rotating (CR) components is indicated by the blue, orange, red and white regions, respectively.  Panel ({\bf b}) shows the average geometric flattening $\langle q \rangle$ as a function of the total angular momentum $\lambda_{\mathrm{total}}$. The horizontal error bars indicate the covered range in $\lambda_z$ ({\bf a}) and  $\lambda_{\mathrm{total}}$ ({\bf b}), whereas the vertical error bars represent the standard deviation. }
\label{fig:zR300}
\end{figure}

\clearpage

\section*{Methods}

We describe below in more detail the orbit-based modeling as well as calibration of systematic biases and uncertainties based on a suite of galaxy simulations.

\paragraph{Stellar Kinematics} 
The stellar kinematics are extracted from the spectral range 3750-4550\,\AA\ for the \emph{CALIFA} V1200 grating, with nominal instrumental resolution of 66\,\kms\ \cite{califakin2017, califadr3}. 
Two-dimensional Voronoi tessellation \cite{Cappellari2003} is used to create spatial bins with a minimum signal-to-noise $S/N = 20$.
The radial extent of the stellar kinematics reaches at least one (SDSS $r$-band) half-light-radius $R_{\mathrm{e}}$ for $95\%$ of the sample, while $77\%$ reach $1.5\,R_{\mathrm{e}}$, and $39\%$ extend beyond $2\,R_{\mathrm{e}}$.

\paragraph{Stellar masses:}
The total stellar mass $M_*$ is not inferred from our best-fitting dynamical model, but instead is derived \cite{Walcher2014} by fitting the spectral energy distributions (SEDs) from multi-band photometry using a linear combination of single stellar population synthetic spectra of different ages and metallicities, adopting a Kroupa \cite{Kroupa2001} IMF.
While our dynamical models yield precise measurements of the total enclosed mass, the separation into stellar, gas and dark matter mass is much less certain due to degeneracies and unavailability of cold gas measurements. 
At the same time, stellar mass obtained through SED fitting is also relatively straightforward to infer from observed and simulated galaxies and thus convenient for comparison of our results with simulated galaxies.

\paragraph{Dynamical models:}
The orbit-based Schwarzschild \cite{Schwarzschild1979} method, which builds galactic models by superposing stellar orbits generated in a gravitational potential, is widely used to model the dynamics of all kind of stellar systems. We assume the system is in steady-state equilibrium with the stars phase-mixed. Rather than single stars evolving with time, an orbit  represents multiple stars on different positions along the orbit at the 'snapshot' we observe the galaxy.

We construct triaxial Schwarzschild models \cite{Schwarzschild1979, vdB2008} of 300 \emph{CALIFA} galaxies in an uniform way as described in a separate paper \cite{Zhu2017a}.  
In brief, the gravitational potential is constructed with a triaxial stellar component embedded in a spherical dark matter halo. 
We first describe a galaxy's image (SDSS r-band image is adopted) by an axisymmetric 2D (Multiple Gaussian Expansion) MGE model \cite{Cappellari2002}, i.e., bar, spiral arms, disk warps or dust lanes are not explicitly fitted\cite{Zhu2017a}. 
Then by adopting a set of viewing angles ($\vartheta, \psi, \phi$) of the galaxy, we de-project the 2D MGE to a triaxial 3D MGE which describes the stellar luminosity \cite{Cappellari2002}. By multiplying a constant stellar mass-to-light ratio $\Upsilon_*$ to the 3D luminosity, we obtain the intrinsic mass density of stars.
The three viewing angles relate directly to $(q, p, u)$ describing the intrinsic shape of the 3D stellar system \cite{Cappellari2002, vdB2008}. We leave $(p,q)$ as free parameters to allow intrinsic triaxial shapes, but fix $u=0.9999$ so that the intrinsic major axis projects onto the projected major axis. While the latter is formally only valid for oblate axisymmetric systems like most galaxies, there is no significant change in the results in case of the mildly triaxial giant elliptical galaxies.
A spherical NFW halo is adopted, with concentration $c$ fixed according to its relation with virial mass $M_{200}$ from cosmological simulations \cite{Dutton2014}. Thus we have four free parameters in the model: the stellar mass-to-light ratio $\Upsilon_*$; $q$ and $p$ describing the intrinsic triaxial shape of the galaxy and a spherical NFW dark matter halo mass $M_{200}$. It is a static potential and figure rotation is not allowed in the model.
The orbit weights are then determined by fitting the orbit-superposition models to the projected and de-projected luminosity density and the two-dimensional line-of-sight stellar kinematics, here the mean velocity and velocity dispersion maps \cite{Falcon-Barroso2016}. 
We do not use regularization in the model so that the orbit weights are independent from each other and fully determined by the least linear $\chi^2$ fitting.

\paragraph{Statistical uncertainties:}
We find these best-fitting models by an iterative search on the four-dimensional parameter space.
In the end we select all the models, across the four-dimensional space, within $\chi^2 - \min(\chi^2) < \sqrt{2N_{\mathrm{obs}}}$. 
In this manner, we find the galaxy's mass distribution, its intrinsic shape, as well as the weights of different orbits that contribute to the best-fitting model. 
We adopt $\sqrt{2N_{\mathrm{obs}}}$ as $1\sigma$ confidence level \cite{Zhu2017a}, where $N_{\mathrm{obs}}$ is the amount of kinematic data constraints. The mean value of orbit distribution of these models, which is usually consistent with that from the single best-fitting model, is taken as the orbit distribution of the galaxy used in Figure~\ref{fig:orbit300_1Re} and Figure~\ref{fig:lzd}, although we only illustrate the orbit distribution of the best-fitting model in Figure 1. The scatter in the orbit distribution among these models is taken as the $1\sigma$ statistical error ($\sigma_{\mathrm{stat}}$). The error bars in Figures 3 are a combination of the latter statistical uncertainties as well as systematic bias and uncertainties which we determine next with the help of galaxy simulations.

\paragraph{Galaxy simulations:}
We evaluate the reliability of our dynamical modeling approach by applying it to simulated galaxies and comparing the resulting orbit distribution to the true orbit distribution directly calculated from the full 6D information of particles in the simulated galaxies.

We use 15 simulated galaxies with different properties (summarized in Supplementary Table 2), 8 massive spiral galaxies with features of warm/cold disk, with/without bars, spiral arms, disk warps or gas, 2 axisymmetric ellipticals (S0s), 2 triaxial elliptical galaxies and another 3 low mass spiral galaxies which are dynamically warmer than the massive spiral galaxies. 
They are from different simulations: 1 is a pure N-body simulation without gas \cite{Shen2010}, while the others are hydrodynamical cosmological simulations, 5 from NIHAO \cite{Wang2015}, 5 from Auriga \cite{Grand2017} and 4 from Illustris \cite{Nelson2015}. Cold gas is included in the galaxies from NIHAO and Auriga \cite{Stinson2015, Marinacci2017}. 

Each simulated galaxy (with long, medium, and short axis x, y, z) is then projected to the observational plane in the following way: we first rotate the galaxy on the plane (x,y) with an angle $\psi$ ($\psi$ only matters when the galaxy is triaxial or has a structure deviating from axisymmetric, e.g. a bar); then we project it to the observation plane with an inclination angle $\vartheta$ (the projected angle between the z axis and line-of-sight); and finally we rotate the projected galaxy to have the projected major axis horizontally aligned.   
We consider projections with different viewing angles ($\vartheta, \psi$) just as the case of \emph{CALIFA} galaxies. For each spiral, we randomly choose a $\psi$ and create 7 mock data sets with inclination angle $\vartheta = 30^o-90^o$. But for $S_1$ we choose three different $\psi$ to test the effect of bar orientation. For each elliptical galaxy, we randomly produce 10 mock data sets with ten pairs of $(\psi, \vartheta)$.  
Finally, we ``observe" each projected galaxy and create mock data with \emph{CALIFA}-like properties \cite{Zhu2017a}.
With these 15 different simulated galaxies, we create 131 mock data sets in total, having data qualities and orientations generally representative of the \emph{CALIFA } sample. 

Each of these 131 mock data set is then taken as an independent galaxy and Schwarzschild models are constructed to fit the mock observations and infer the best-fit orbit distribution. 
Next, we calculate the true orbit distribution from the particles with known full 6D information. 
For the simulated spiral galaxies, we use a single snapshot and select those particles which are close in energy $E$, angular momentum $L_z$ and the total angular momentum amplitude $L$. Under the assumption that these particles are on similar orbits in an axisymmetric system, we then compute the corresponding phase-space averages of $r$ and $\lambda_z$.
For the elliptical galaxies that are triaxial, we instead take a time-average approach. We take single-snapshot positions of all particles to calculate the corresponding smooth gravitational potential with a tree code \cite{Barnes1986}. Within this 'frozen' gravitational potential, we then compute for each particle its orbit starting from the 6D position-velocity values of the particle in the snapshot. After $\sim$200 orbital periods, we then calculate the time-averaged radius $r$ and circularity $\lambda_z$.
We checked for two near axisymmeric simulated galaxies that the phase-space average yields a very similar orbit distribution $p(\lambda_z, r)$ as the time average method which is computationally costly.

For each of the 131 mock data sets, the orbit distributions $p(\lambda_z, r)$ obtained by our models generally match well the true orbit distribution of the simulated galaxies (Supplementary Figure 1 and Figure 2). 
To quantitatively evaluate the uncertainties of our results, we integrate $p(\lambda_z, r)$ over $r<R_{\rm e}$ and obtain $p(\lambda_z)$ within $R_{\rm e}$.  
We then further divide the orbit distribution $p(\lambda_z)$ into cold, warm, hot and CR components and calculate the orbit fraction of each component ($f_{\mathrm{cold}}, f_{\mathrm{warm}}, f_{\mathrm{hot}}, f_{\mathrm{CR}}$). 
The model recovered orbit fraction of four components ($f_{\mathrm{cold}}, f_{\mathrm{warm}}, f_{\mathrm{hot}}, f_{\mathrm{CR}}$) generally matches the true values one-to-one (Supplementary Figure 3). 

\paragraph{Systematic biases and uncertainties:}
The 15 simulated galaxies include galaxies with different features and Hubble types. To summarize the model uncertainties for different types of galaxies, we define the relative deviation $d = {(f_{\mathrm{model}} - f_{\mathrm{true}}) / f_{\mathrm{model}}}$ for each single mock data set.
The average and standard variations of $d$ are calculated with $\overline{d} =$ $ {\overline{(f_{\mathrm{model}} - f_{\mathrm{true}})} / \overline{f_{\mathrm{model}}}}$ and $\sigma(d) = {\sigma(f_{\mathrm{model}} - f_{\mathrm{true}}) / \overline{f_{\mathrm{model}}}}$, for each type of galaxy.
We consider $\overline{d_{\mathrm{cold}}} , \, \overline{d_{\mathrm{warm}}} , \,  \overline{d_{\mathrm{hot}}} , \, \overline{d_{\mathrm{CR}}} $ as the relative systematic model bias,
and $\sigma(d_{\mathrm{cold}})$, $\sigma(d_{\mathrm{warm}})$ , $\sigma(d_{\mathrm{hot}})$ , $\sigma(d_{\mathrm{CR}})$ as the relative systematic error to be added to the statistical uncertainty for each galaxy of a given type.

We do not find a clear relation of the deviations $d$ with the complicated features of spiral galaxies; bars, spiral arms, warps or gaseous disk do not seem to bias our recovery of the whole orbit distributions (Supplementary Figure 4). 
We divide the galaxies into four groups: 3 low-mass spiral galaxies (with total stellar $M_* \lesssim 10^{10} \, M_{\odot}$), 8 high-mass spiral galaxies, 2 near-axisymmetric lenticular galaxies, and 2 triaxial elliptical galaxies.
For each group of galaxies, we calculate the average $\overline{d}$ and standard deviation $\sigma(d)$ for the cold, warm, hot and CR components as shown in Supplementary Table 3; we consider them representative of the relative systemic bias and relative systemic errors of the corresponding types of \emph{CALIFA} galaxies. 
The model derived orbit distributions of the three simulated low-mass spiral galaxies are similar to those of the low-mass \emph{CALIFA} galaxies.

The fit is uncertain for simulated galaxy g2.42e11, which shows an off-center non-symmetric maximum velocity dispersion, which may or may not be physical, but can not be fitted irrespectively.
For very face-on spiral galaxies with $\vartheta < 30^o$, the uncertainty caused by de-projection is higher; we have a $\sim 50\%$ chance to significantly over-estimate the cold and CR component fraction, while under-estimating the warm and hot components. In the \emph{CALIFA} sample, we have only two spiral galaxies with $\vartheta < 20^o$ and another five with $20^o<\vartheta < 30^o$. The stellar orbit distributions of these seven galaxies do not show a typical biased pattern and there is no noticeable difference in our results by excluding them. 

\paragraph{Overall uncertainties:} In the end, the uncertainties of orbit fractions shown in Figure 3 include statistical errors, systematic biases and systemic errors. 
The scatter of orbit fractions in the models within $1 \sigma$ confidence level estimated before are used as statistical errors with typically $\sigma_{\mathrm{stat}} \sim 0.2f_{\mathrm{model}}$.
We obtained systematic biases $\mathcal{D} =\overline{d}f_{\mathrm{model}}$ and systematic errors $\sigma_{\mathrm{sys}} = \sigma(d)f_{\mathrm{model}}$ from the above tests representing a typical low-mass spiral, high-mass spiral, lenticular or elliptical galaxy. 
There are $n \sim 30$ galaxies in each bin in Figure~3. 
We then first take equal length for the lower and upper error bar $\sigma = \sqrt{\overline{\sigma_i^2}/n}$ with $\sigma_i^2 = \sigma_{\mathrm{stat}}^2 + \sigma_{\mathrm{sys}}^2$ for each galaxy. Then we include the systematic bias $\mathcal{D} = \overline{\mathcal{D}_i}$ to the upper or lower error bar according to its sign. For instance, if $\mathcal{D}<0$, it means the model under-estimates the value, thus we increase the upper error bar by $|\mathcal{D}|$ to cover the true value.  


\paragraph{Caveats:}

In the final sample of 250 galaxies, there are 69 galaxies with strong bars, 52 with weak bars and 129 with no bar, classified by eye.
Generally, we find that the barred and unbarred galaxies have similar orbit distributions as a function of total stellar mass, although galaxies with strong bars have a slightly higher fraction of cold orbits.
This, however, is a result of bars being more frequent in late-type galaxies with cold disks and less likely in early-type galaxies.

We select 31 galaxies with dust lanes by eye within the 269 isolated galaxies with $R_{\mathrm{max}} > R_{\rm e}$.
We find that for the 12 near edge-on galaxies (with inclination angle $\vartheta > 70^o$), their orbit distributions are comparable to the dust-free galaxies at the same inclination angle regions and same mass ranges. 
The 19 more face-on dusty galaxies have significant lower fraction of cold orbits and more CR orbits compared to the dust-free galaxies of similar inclination angles and masses. 
We presume the orbit distribution of the 19 dusty (more face-on) galaxies are biased by our modelling process which does not include the effect of dust.  
It is hard to either correct the effects of dust in our model or evaluate the effect by testing with simulated galaxies. Thus we exclude the 19 galaxies from the sample, and keep only the remaining 250 galaxies for further consideration. 

There are some spuriously high velocity dispersion bins in the outer region of \emph{CALIFA} data due to the limited spectral resolution \cite{Falcon-Barroso2016}. 
We created new kinematic maps by removing the points with $S/N <20$ but simultaneously $d\sigma/\sigma<0.2$; they usually have a significantly higher dispersions than other bins in the galaxy. Removing these points makes a notable difference in the kinematic map for 50 galaxies, which are mostly late-type spirals. 
We run the models with the original kinematic maps (no data points excluded) and with the new kinematic maps (points excluded), respectively, for these 50 galaxies. 
Nearly $90\%$ of the spuriously high dispersion points are in outer regions $r>R_{\rm e}$ of galaxies. The orbital distribution within $R_{\rm e}$ are only mildly affected by these points and do not affect the inferred orbital distribution. 
For the 50 selected galaxies, we take the orbital distribution from the best-fitting models constrained by the data with these points excluded. 

Due to the limited data coverage, we only present the orbit distribution within $R_{\rm e}$.  The over-all orbit distribution of a galaxy could be different in two aspects: first the near-circular cold orbits may be more prominent in the outer regions, and second we use the time-averaged radius $r$ to describe each orbit, therefore, in particular low-angular momentum orbits at their apocenters extend well beyond $r$, and hence $R_{\rm e}$. We need kinematic maps extending to larger radii to overcome these caveats. However, the present results could be used for direct comparison  with the inner regions of simulated galaxies.

\paragraph{Data Availability Statement}
The data of Figure 3 are included in the Supplementary file. Other related data that support this study are available from the corresponding author upon reasonable request.

\paragraph{Additional References}

\newpage

\section*{Supplementary}

Supplementary Table~1 lists the values and asymmetric error bars as presented in Figure~3 of the main text. 
Supplementary Table~2 presents an overview of the galaxy simulations used to test our orbit-based modelling approach and to calibrate the systematic biases and uncertainties.
Supplementary Table~3 gives the latter systematic biases and uncertainties which in turn have been included in the asymmetric error bars in Figure~3 of the main text.
Supplementary Figures~1--4 show how our modelling approach recovers well the true orbit distributions of the simulated galaxies. 

\begin{table*}
\caption{{\bf Supplementary Table~1: } Volume-corrected average luminosity fractions (SDSS r-band) of cold, warm, hot and CR components in eight bins within different ranges of stellar mass $M_{\mathrm{star}}$ (as plotted in Figure 3 in the main paper).  The sub/super scripts in the top row denote the range of $\log M_{\mathrm{star}}$ per bin while in the remaining rows they denote the $1\sigma$  lower/upper uncertainty on the fractions of the cold, warm, hot and counter-rotating (CR) orbital components.}
\label{tab:lzd}\footnotesize
\begin{tabular}{*{9}{l}}
\hline
\hline
$\log{ M_{\mathrm{star}} }$   &  $9.6^{0.4}_{-0.9}$   &  $10.2^{0.1}_{-0.2}$  &   $10.4^{0.1}_{-0.1}$   &  $10.6^{0.01}_{-0.1}$ &  $10.8^{0.01}_{-0.01}$  & $10.97^{0.08}_{-0.07}$   & $10.2^{0.2}_{-0.15}$ &  $11.6^{0.34}_{-0.2}$  \\
\hline
$f_{\mathrm{cold}}$    &  $0.20_{-0.04}^{0.02}$  &  $0.35_{-0.03}^{0.03}$  &  $0.28_{-0.02}^{0.02}$ & $0.19_{-0.02}^{0.02}$  & $0.20_{-0.02}^{0.03}$  & $0.16_{-0.01}^{0.03}$  & $0.17_{-0.02}^{0.03}$  & $0.07_{-0.01}^{0.03}$\\

$f_{\mathrm{warm}}$ & $0.40_{-0.04}^{0.03}$  & $0.38_{-0.02}^{0.03}$  & $0.38_{-0.02}^{0.03}$  & $0.44_{-0.04}^{0.03} $ & $0.44_{-0.03}^{0.03}$  & $0.40_{-0.06}^{0.03}$  & $0.34_{-0.06}^{0.02}$  & $0.26_{-0.12}^{0.02}$\\

$f_{\mathrm{hot}}$      &  $0.23_{-0.02}^{0.18}$   &  $0.19_{-0.02}^{0.06}$  & $0.24_{-0.02}^{0.07}$  & $0.28_{-0.02}^{0.08}$  & $0.29_{-0.02}^{0.06}$  & $0.35_{-0.02}^{0.08}$  & $0.38_{-0.03}^{0.10}$ & $0.51_{-0.03}^{0.18}$   \\

$f_{\mathrm{CR}}$     & $0.17_{-0.08}^{0.01}$    &  $0.08_{-0.04}^{0.01}$  & $0.09_{-0.04}^{0.01}$  & $0.08_{-0.03}^{0.01}$ & $0.06_{-0.02}^{0.01}$  & $0.09_{-0.03}^{0.01}$ &
$0.1_{-0.04}^{0.01}$ & $0.15_{-0.08}^{0.01}$ \\
  \hline
 \end{tabular}
\end{table*}

\begin{table}
\caption{{\bf Supplementary Table~2:} The 15 simulated galaxies used to test our orbit-based modelling approach and to calibrate the systematic biases and uncertainties. From left to right, the galaxy name, stellar mass $M_\star$, neutral hydrogen mass $M_{\mathrm{HI}}$, source of the simulations, Hubble types, specific properties, the number of projections taken to create mock data sets, and ranges of the two viewing angles $\vartheta$ and $\psi$ (both in degrees). Spiral galaxies are divided into low-mass spirals (Low-S) and high-mass spirals (High-S).  A total of 131 mock data sets are created.}
\scriptsize\centering
\begin{tabular}{*{9}{l}}
\hline
\hline
Name  & $M_\star/M_\odot$ & $M_{\mathrm{HI}}/M_\odot$ & source &  type & property & \# proj.\ & $\vartheta(^\circ)$ &  $\psi(^\circ)$ \\
\hline
$S_1$ & $4.25e10$       & no gas& N-body & High-S & warm disk, strong bar  & 21 & $30^o-90^o$ & $\psi_{\mathrm{bar}} = 0^o, 45^o, 90^o$ \\
$g2.57e11$ & $1.0e10$   & $6e9$ &NIHAO & Low-S  & warm disk & 7 & $30 - 90$ & random\\
$g2.42e11$ & $5.36e9$   & $3e9$ &NIHAO & Low-S & warm disk  & 7 & $30 - 90$ & random\\
$g5.02e11$ & $1.46e10$  & $1.5e10$ & NIHAO & Low-S & warm disk & 7 &  $30 - 90$ & random\\
$g7.55e11$ & $3.03e10$   &$3.0e10$  & NIHAO & High-S & warm disk & 7 &  $30 - 90$ & random\\
$g8.26e11$ & $4.52e10$   & $2.0e10$ &NIHAO & High-S & warm disk  & 7 & $30 - 90$& random\\
$Au-25$  &$ 3.14e10$   & $1.56e10$  & Auriga & High-S & cold disk, warps, weak bar & 7 & $30 - 90$ & random\\
$Au-6$  &$ 4.75e10$   & $1.5e10$ & Auriga & High-S & spiral arms, weak bar  & 7 & $30 - 90$ & random \\
$Au-5$  &$ 6.7e10$   & $7.2e9$ & Auriga & High-S & spiral arms, weak bar & 7 & $30 - 90$ & random\\
$Au-4$  &$ 7.1e10$   & $7.3e9$ &Auriga & High-S & warm disk, disturb & 7 & $30 - 90$ & random\\
$Au-23$  &$ 9.02e10$ & $1.45e10$ & Auriga & High-S & warps, strong bar & 7 & $30 - 90$ & random \\
sh110569 &      $3.69e11$      & -    & Illustris & S0 & oblate & 10 & random & random \\
sh202062 &      $3.28e11$      & -    &Illustris & S0 & oblate  & 10 & random & random\\
sh16942  &      $2.92e11$      & -       & Illustris & E & triaxial & 10 & random & random\\
sh312924 &      $3.12e11$     & -      & Illustris & E & triaxial  & 10 & random & random\\
  \hline
 \end{tabular}
\end{table}

\begin{table}
\caption{{\bf Supplementary Table~3:} We divide the 15 simulated galaxies into four groups: 3 low-mass spiral galaxies, 8 high-mass spiral galaxies, 2 lenticular galaxies, and 2 triaxial elliptical galaxies. 
For each group, $\overline{d}$ and $\sigma(d)$ are the average and standard deviation of the residual fraction $d = (f_{\mathrm{model}}- f_{\mathrm{true}}) /f_{\mathrm{model}}$ of the cold, warm, hot and counter-rotating (CR) orbital components. The resulting, $\overline{d}$ and $\sigma(d)$ are adopted as the typical relative systemic bias and relative systemic uncertainties of the four components for the corresponding type of \emph{CALIFA} galaxies. }
\scriptsize\centering
\begin{tabular}{*{9}{l}}
\hline
\hline
Type  & $\overline{d_{\mathrm{cold}}}$ & $\overline{d_{\mathrm{warm}}}$   &  $\overline{d_{\mathrm{hot}}}$ & $\overline{d_{\mathrm{CR}}}$ & $\sigma(d_{\mathrm{cold}})$ &  $\sigma(d_{\mathrm{warm}})$ & $\sigma(d_{\mathrm{hot}})$ & $\sigma(d_{\mathrm{CR}})$ \\
\hline
Low-S & 0.10 & 0.03 & -0.71 & 0.36 & 0.41 & 0.17 & 0.20 &0.19\\
High-S & 0.02 &  -0.01 & -0.21 &  0.31 &  0.33 & 0.17 & 0.29 & 0.40 \\
S0 & -0.25 &  0.07 & 0.11 & -0.04 & 0.30 & 0.17 & 0.11 & 0.27\\
E & -0.27 & 0.48 & -0.36 & 0.49 & 0.62 & 0.25 & 0.15 & 0.23 \\ 
  \hline
 \end{tabular}
\end{table}

\begin{figure}
\centering\includegraphics[width=12cm]{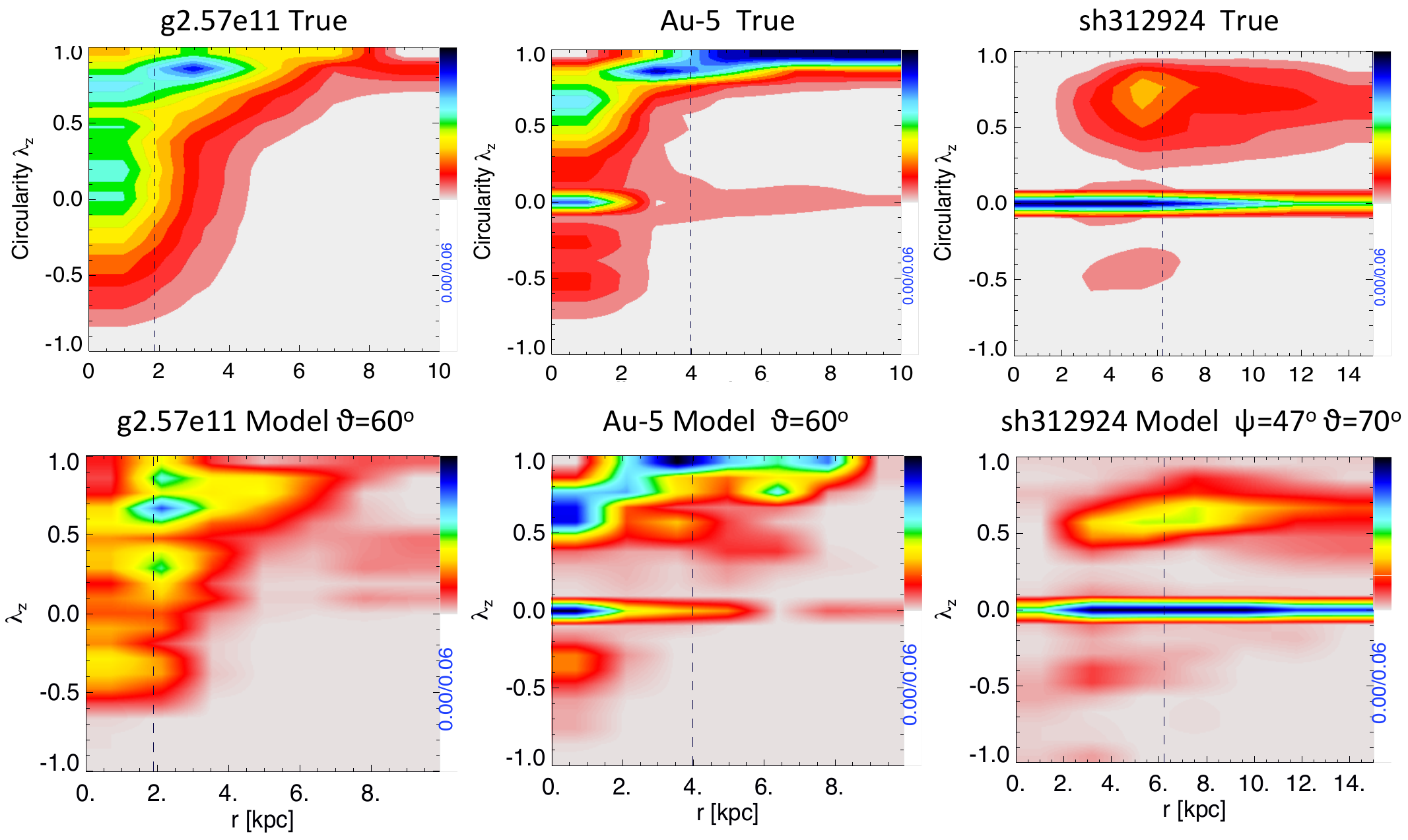}
\caption{{\bf Supplementary Figure~1:} Recovery of the orbital distribution $p(\lambda_z, r)$ as function of circularity $\lambda_z$ and intrinsic radius $r$. 
{\bf Top:} The true orbital distribution calculated from stellar particles with full 6D information in the simulated galaxies, g2.57e11 (left), Au-5 (middle) and sh312924 (right). 
{\bf Bottom:} The orbital distribution from our best-fitting model constrained by mock data at viewing angles $\vartheta = 60^o$ for g2.57e11 and Au-5, and $(\psi, \vartheta) = (47^\circ, 70^\circ)$ for sh312924.}
\label{fig:mock_orbit2d}
\end{figure}

\begin{figure}
\centering\includegraphics[width=17cm]{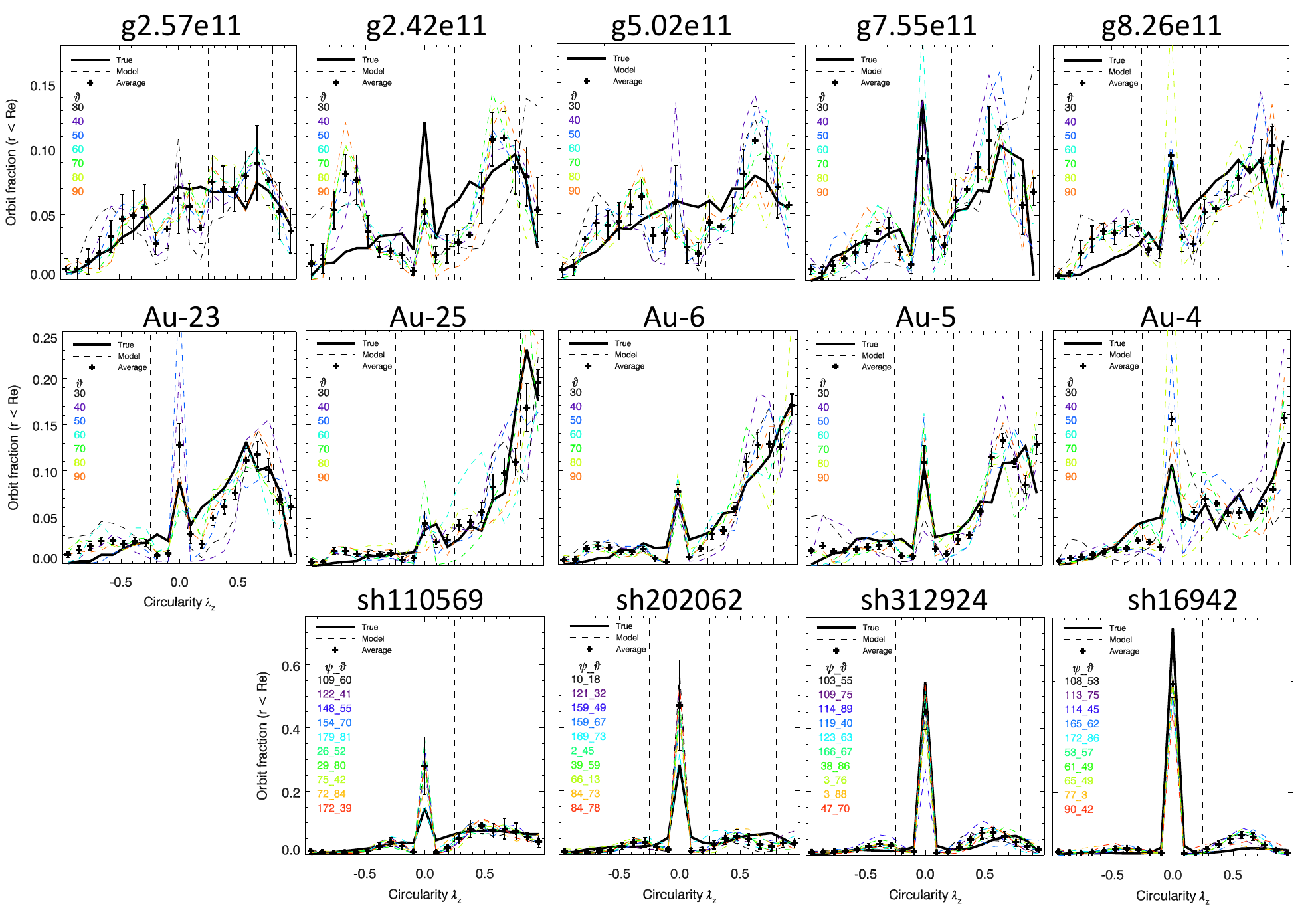}
\caption{{\bf Supplementary Figure~2:} Orbital circularity $\lambda_z$ distribution within $R_e$ for 14 of the 15 simulated galaxies (similar figure for $S_1$ in Zhu et al.\ 2017a). 
In each panel, the true $\lambda_z$ distribution of the simulated galaxy is shown as the black solid line. 
Each coloured line is the average orbital distribution from modelling a mock data set with viewing angles as labeled.
The error bars show the typical $1\sigma$ scatter for each mock data set. 
Crosses denote the average of the seven/ten different dashed lines for spiral/elliptical galaxies.}
\label{fig:mock_orbit1d}
\end{figure}

\begin{figure}
\centering\includegraphics[width=14cm]{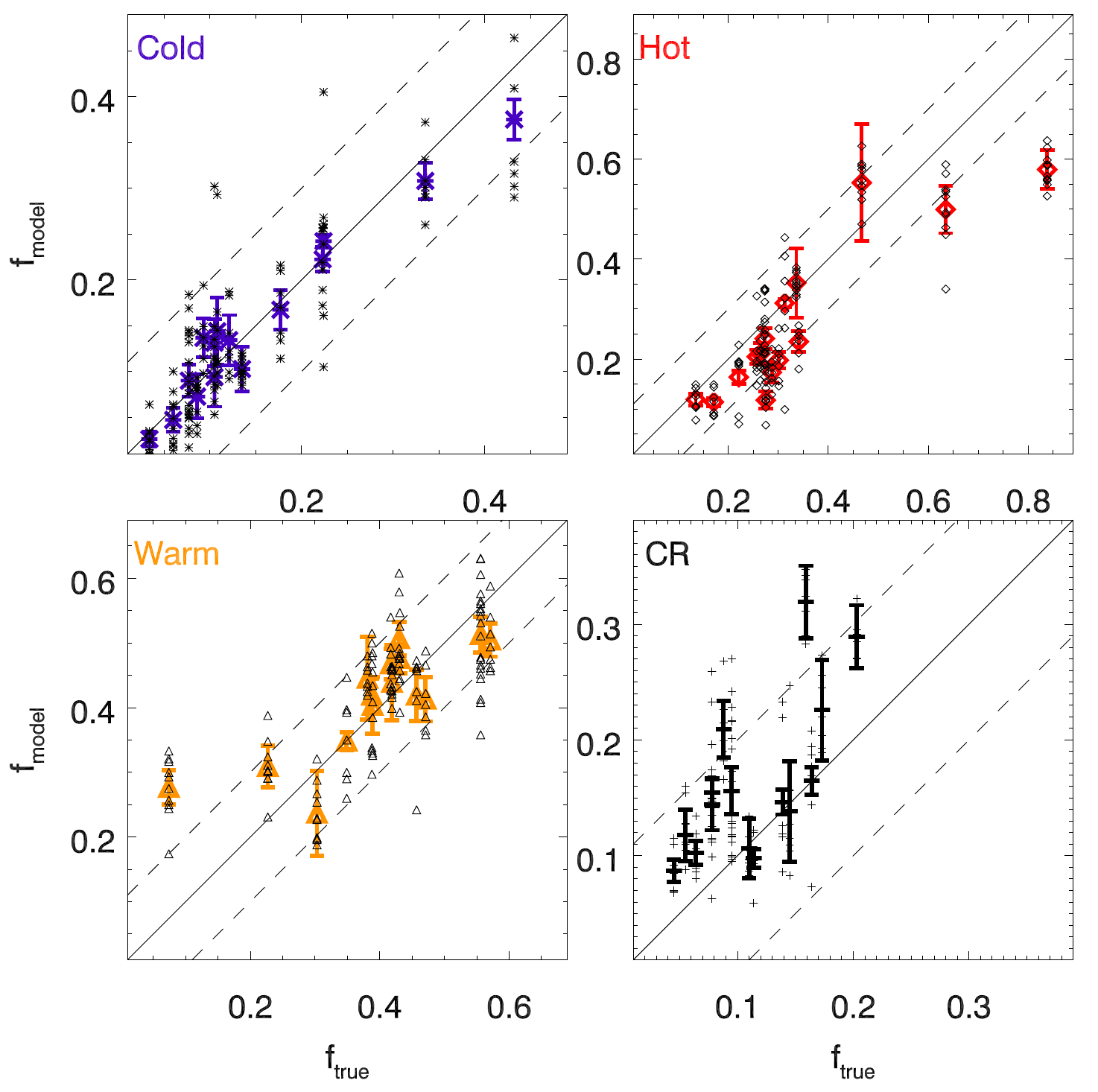}
\caption{{\bf Supplementary Figure~3:} One-to-one comparison of the true value and model recovered cold, warm, hot and CR orbit fractions within $R_e$.  Each thin symbol represent 
one mock data set, while each thick coloured symbol represent the average from the seven/ten mock data sets with different viewing angles for spiral/elliptical galaxies. 
The error bars are the typical $1\sigma$ scatter of each single set of model (each thin symbol). The solid line is the one-to-one line, the two dashed lines deviate with $\pm0.1$ from the solid line.}
\label{fig:mock_t1}
\end{figure}

\begin{figure}
\centering\includegraphics[width=8.6cm]{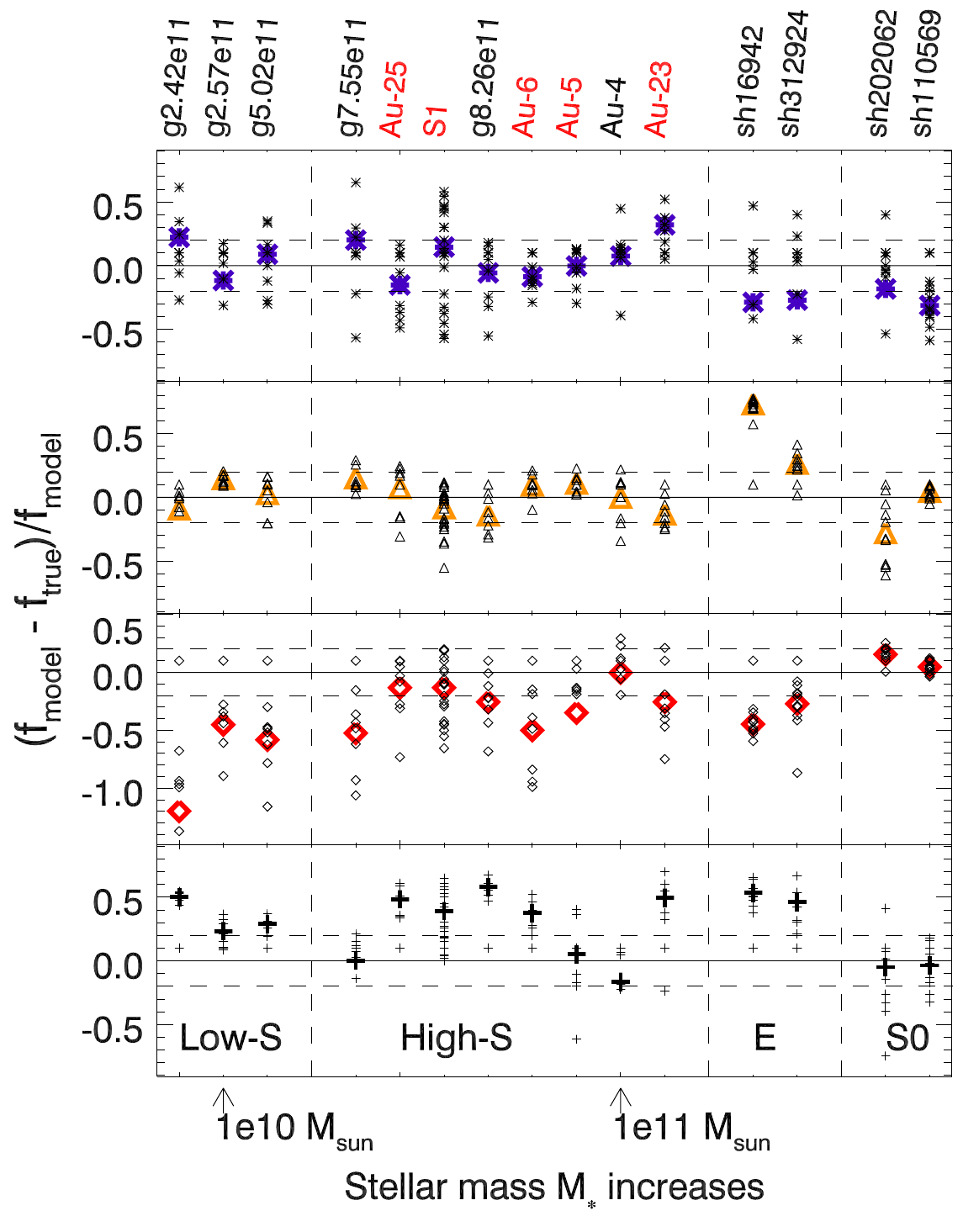}
\caption{{\bf Supplementary Figure~4:} Relative deviations of cold, warm, hot and CR orbit fractions from top to bottom, symbols the same as in Figure~\ref{fig:mock_t1}.  
The 15 galaxies are sorted according to increasing total stellar mass with names in red being barred galaxies. 
For the eight high-mass spirals, 5 with a bar and 3 without a bar, we do not see a significantly different/larger deviations of any components that could be caused by a bar. 
}
\label{fig:mock_t2}
\end{figure}

\end{document}